\documentclass[12pt]{article}
\begin{document}
\input{epsf}
\Large
\begin{center}   
\bf{Study of the multiplicity of hadrons on nuclei}
\end{center}

\normalsize
\begin{center}
      L.~Grigoryan\footnote{e-mail: leva@mail.desy.de}
\end{center}

\begin{center}
Alikhanyan National Science Laboratory
      (Yerevan Physics Institute), Br.Alikhanian 2, 375036 Yerevan, Armenia
\end{center}

\begin {abstract}
\hspace*{1em}
The improved two-scale model with symmetric Lund distribution function
is used to perform the fit to the semi-inclusive deep-inelastic scattering
(SIDIS) data on nuclear targets of HERMES experiment at DESY. The ratio of
hadron multiplicity on nuclear target to the deuterium one is chosen as
observable, as usually. Some peculiarities of the fit are discussed.
The two-parameter's fit gives satisfactory agreement with the data. The
results of the present and previous fits are compared.
\end {abstract}
\twocolumn  
\section{Introduction}
\normalsize 
\hspace*{1em}
Hadronic reactions in a nuclear medium can shed additional light on the
hadronization process. In comparison with other reactions electroproduction
has the virtue that energy and momentum of the struck parton are well
determined, as they are tagged by the scattered lepton. Study of hadron
production in SIDIS on nuclear targets offers an opportunity to
investigate the quark (string, color dipole) propagation in nuclear
matter and the space-time evolution of the hadronization process.
For this purpose we investigate the nuclear attenuation (NA), which is a
ratio of differential hadron multiplicity on a nucleus to that on
deuterium:\\
\begin{eqnarray}
\nonumber
R_{M}^{h}(\nu,z) = \frac{\Big(\frac{N^{h}(\nu,z)}{N^{e}(\nu)} \Big)_{A}}
{\Big(\frac{N^{h}(\nu,z)}{N^{e}(\nu)} \Big)_{D}} \hspace{0.2cm},
\end{eqnarray}
where $z = E_h/\nu$, $E_h$ and $\nu$ are energies of the final hadron and
virtual photon, $N^{h}(\nu,z)$ is the number of semi-inclusive hadrons at
given $\nu$ and $z$ and $N^{e}(\nu)$ is the number of inclusive DIS
leptons at given $\nu$. Subscripts $A$ ($D$) denote that reaction takes
place on nucleus (deuterium). In the above formula more variables like -$Q^2$
(the photon virtuality) and $p_t$ (hadron transverse momentum) over which
the NA is averaged, are not written. The simple version of the string model,
so called two-scale model (TSM), was introduced in Ref.~\cite{ashman}. In
Ref.~\cite{akopov2} improved version of TSM (ITSM) was proposed. In our
previous work~\cite{akopov7} ITSM was used for a fitting of the
SIDIS data of HERMES experiment on nuclear targets~\cite{airapet1}.
For the fit we used those versions of the distribution function, which
possible to present in the finite form. In result, two versions were used:
(i) the leading hadron distribution function, which corresponds to the term
with rank equal one of the distribution function; and (ii) the standard
Lund distribution function, which possible to sum analytically over all
ranks of hadrons carrying momentum $z$. Unfortunately, more realistic symmetric
Lund distribution function it is not possible to sum over all ranks
analytically. For this reason it was not included in the fit.\\
\hspace*{1em}
The aim of this work is a fitting similar to that which was done in 
Ref.~\cite{akopov7} with using the symmetric Lund distribution function
and a comparison of the results of fit with the results of the previous one.\\
\hspace*{1em}
The paper is organized as follows. In section 2 some
details of the fitting procedure with symmetric Lund distribution function are 
discussed. In section 3 we compare results of the present fit with the previous
one and discuss them. Conclusions are given in section 4.
\section{Some features of fit}
\normalsize
\hspace*{1em}
In the string models, for the construction of fragmentation functions, the scaling
functions $f(z)$ are introduced.
For calculations we use~\cite{andersson,sjostrand}:\\
(i) standard Lund scaling function
\begin{eqnarray}
\nonumber
f(z) = (1 + C)(1 - z)^{C} ,
\end{eqnarray}
where $C$ is the parameter which controls the steepness of the standard Lund
fragmentation function ($C = 0.3$);\\
(ii) symmetric Lund scaling function
\begin{eqnarray}
\nonumber
f(z) = Nz^{-1}(1 - z)^{a}exp(-bm_{\perp}^{2}/z) ,
\end{eqnarray}
where $a$ and $b$ are parameters of the model ($a=0.3$, $b=0.58 GeV^{-2}$),
$m_{\perp}=\sqrt{m_{h}^{2}+p_{\perp}^{2}}$
is the transverse mass of final hadron, $N$ is normalization factor.
We begin by considering the distribution of the constituent formation lengths $l$
of hadrons carrying fractional energy $z$. The simplified version without
identifying flavor of partons has the form:
\begin{eqnarray}
\nonumber
D_{c}^{h}(L,z,l) = \Big(f(z)\delta(l-L+zL)+
\end{eqnarray}
\begin{eqnarray}
\nonumber
\sum_{i=2}^{n}D_{ci}^{h}(L,z,l)\Big)\theta(l)\theta(L-zL-l)
\hspace{0.3cm}  ,
\end{eqnarray}
where $L = \nu/\kappa$ is the full hadronization length, $\kappa$ is the string
tension ($\kappa$ = 1 GeV/fm). The functions $D_{ci}^{h}(L,z,l)$ are distributions
of the constituent formation length $l$ of the rank $i$ hadrons carrying fractional
energy $z$.
\begin{figure}[!h]
\begin{center}
\epsfxsize=7.cm
\epsfbox{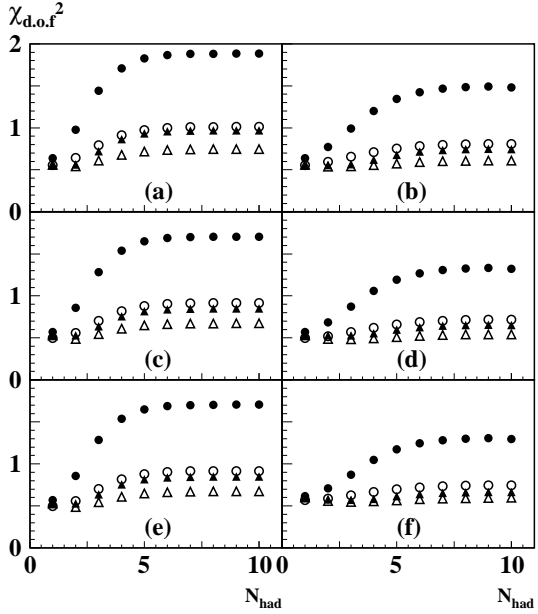}
\end{center}
\caption{\label{fig:fig1}
{
$\hat{\chi}^2$ as a function of $N_{had}$. Left part
presents results of fit with standard Lund distribution function and
right part results of fit with symmetric
Lund distribution function. Panels (a), (b); (c), (d); (e), (f) correspond NDF 1;
NDF 2; NDF 3, respectively.
Filled circles, open circles,
filled triangles, open triangles are results of fit with $\sigma^{str}$
corresponging eqs.(7), (8), (9), (10) of Ref.~\cite{akopov7}.
}}
\end{figure}
For calculations we use the average value of the formation length $\tau_c=<l>$:
\begin{eqnarray}
\nonumber
\tau_c = \int_0^{\infty} ldlD_{c}^{h}(L,z,l)/\int_0^{\infty} dlD_{c}^{h}(L,z,l)
\hspace{0.2cm}.
\end{eqnarray}
The same versions of string-nucleon cross sections, $\sigma^{str}$,
and the same set of the nuclear density functions (NDF) as in Ref.~\cite{akopov7}
are used for fit. The semi-inclusive data~\cite{airapet1} of HERMES experiment
on four nuclear targets (helium, neon, krypton, xenon) and deuterium are used.
As in previous fit we use one and two dimensional data
\footnote{The one (two) dimensional data means that data are presented as
functions of one (two) variables.}
for pions, all together 332 experimental points.
\begin{table}[!h]
\begin{tabular}{|c|c|c|c|}
\hline
\multicolumn{4}{|c|}{$\sigma^{str}(7)$} \\
\hline
$NDF$ & $\sigma_{q} \hspace{0.2cm} mb$ & $c$ & $\hat{\chi}^2$ \\
\hline
1 & $0.00 \pm 0.01$ & $0.365 \pm 0.013$ & 1.48 \\
\hline
2 & $0.00 \pm 0.01$ & $0.336 \pm 0.011$ & 1.32 \\
\hline
3 & $0.00 \pm 0.01$ & $0.307 \pm 0.012$ & 1.30 \\
\hline
\multicolumn{4}{|c|}{$\sigma^{str}(8)$} \\
\hline
$NDF$ & $\sigma_{q} \hspace{0.2cm} mb$ & $c$ & $\hat{\chi}^2$ \\
\hline
1 & $1.74 \pm 0.16$ & $0.172 \pm 0.014$ & 0.81 \\
\hline
2 & $1.88 \pm 0.15$ & $0.163 \pm 0.011$ & 0.71 \\
\hline
3 & $1.84 \pm 0.15$ & $0.140 \pm 0.009$ & 0.74 \\
\hline
\multicolumn{4}{|c|}{$\sigma^{str}(9)$} \\
\hline
$NDF$ & $\sigma_{q} \hspace{0.2cm} mb$ & $c$ & $\hat{\chi}^2$ \\
\hline
1 & $0.00 \pm 0.01$ & $0.112 \pm 0.009$ & 0.75 \\
\hline
2 & $0.00 \pm 0.02$ & $0.088 \pm 0.009$ & 0.66 \\
\hline
3 & $0.00 \pm 0.02$ & $0.059 \pm 0.009$ & 0.66 \\
\hline
\multicolumn{4}{|c|}{$\sigma^{str}(10)$} \\
\hline
$NDF$ & $\sigma_{q} \hspace{0.2cm} mb$ & $c$ & $\hat{\chi}^2$ \\
\hline
1 & $2.17 \pm 0.13$ & $0.098 \pm 0.010$ & 0.61 \\
\hline
2 & $2.34 \pm 0.14$ & $0.088 \pm 0.010$ & 0.54 \\
\hline
3 & $2.43 \pm 0.15$ & $0.069 \pm 0.010$ & 0.59 \\
\hline
\end{tabular}
\vskip 0.5cm
\caption{\label{tab:tab1} {Values of fitting parameters and
$\hat{\chi}^{2}$ in case of Symmetric Lund distribution function
and total errors. Numbers in parentheses indicate the corresponding
equations from the Ref.~\cite{akopov7}.}}
\end{table}
The fit was performed to tune two parameters: the initial value of
string-nucleon cross section $\sigma_{q}$ and coefficient $c$.\\
\hspace*{1em}
The quantitative criterium $\hat{\chi}^2$ was used:
\begin{eqnarray}
\nonumber
\hat{\chi}^2 = \frac{1}{(n_{exp}-n_{par}-1)}\times
\end{eqnarray}
\begin{eqnarray}
\nonumber
\sum_{n=1}^{n_{exp}}
\Big(\frac{R_{M}^{h}(theor)-R_{M}^{h}(exp)}{\Delta R_{M}^{h}(exp)} \Big)^{2} 
\hspace{0.2cm},
\end{eqnarray}
where $n_{exp}$ and $n_{par}$ are numbers of experimental points and parameters;
$R_{M}^{h}(theor)$ is the theoretical value for ratio at given point;
$R_{M}^{h}(exp)$ and $\Delta R_{M}^{h}(exp)$ are experimental value of
$R_{M}^{h}$ and its uncertainty at given point.\\
\hspace*{1em}
Let's consider the features of fitting with symmetric Lund scaling function.
The corresponding distribution function it is impossible to analytically sum
over all ranks of hadrons as in case of standard Lund scaling function,
therefore we must restrict ourself by finite sum over ranks of hadrons.
The recursion equation from Ref.~\cite{bialas2} is used for calculation of
the distribution functions. We performed fits with the sums of ranks until
$n=N_{had}$ and convinced, that beginning from $N_{had}=5$ the values of
parameters and $\hat{\chi}^2$ do not change essentially. The close result is
obtained for the case of standard Lund scaling function. In this case the
results for finite sums for $N_{had} \ge 5$ practically coincide with the result
for infinite sum. Results of fits are presented in Fig. 1. In the left part of
Fig. 1 (panels (a), (c), (e)) the results of fit with standard Lund distribution
function and in the right part of Fig. 1 (panels (b), (d), (f)) the results of
fit with symmetric Lund distribution function are presented. Panels (a), (b);
(c), (d); (e), (f) correspond NDF 1; NDF 2; NDF 3, respectively. We would like
to remind that NDF 1; NDF 2; NDF 3, correspond eqs. (14); (15); (16) of
Ref.~\cite{akopov7}. Filled circles, open circles, filled triangles, open
triangles are results of fit with $\sigma^{str}$ corresponging eqs.(7), (8),
(9), (10) of Ref.~\cite{akopov7}.
\section{Comparison with data and discussion}
\hspace*{1em}
The results of fit with symmetric Lund distribution
function and $N_{had}=10$ are presented in the Table~\ref{tab:tab1}.
\begin{figure}[!h]
\begin{center} 
\epsfxsize=7.cm
\epsfbox{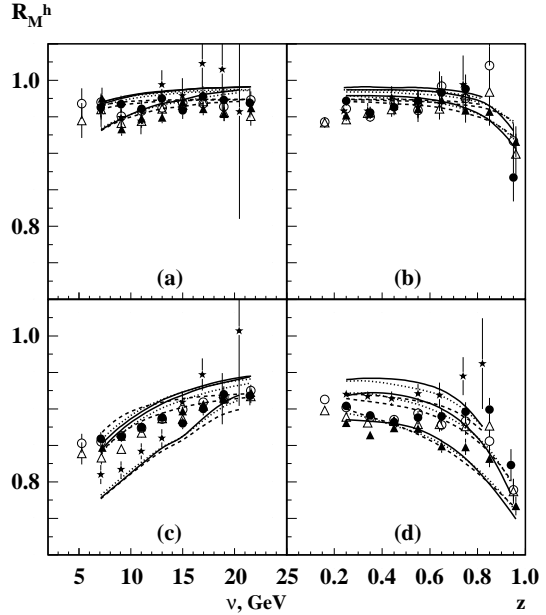}
\end{center}
\caption{\label{fig:fig2} {The two dimensional data.
The ratio $R_{M}^{h}$ for charged pions on $^{4}He$ (panels a, b) and
$^{20}Ne$ (c, d) nuclei as a function of
$\nu$ (left panels) and $z$ (right panels).
Experimental points from Refs.~\cite{airapet1},~\cite{airapet2}.
}}
\end{figure}  
\begin{figure}[!h]
\begin{center}
\epsfxsize=7.cm
\epsfbox{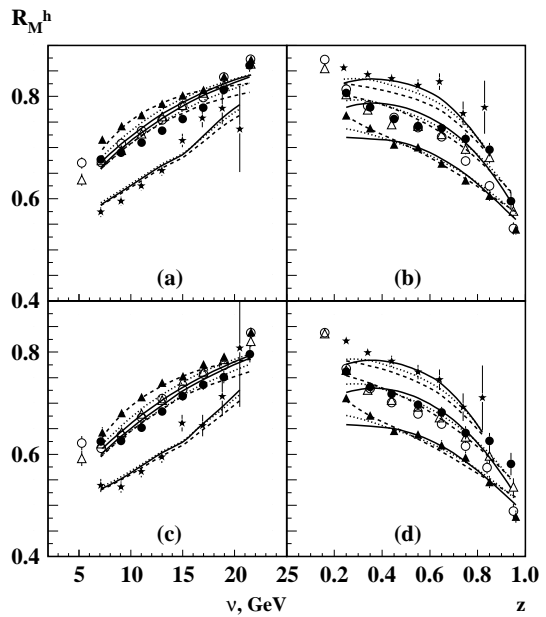}
\end{center}
\caption{\label{fig:fig3}{The same as described in the caption of the
Fig.2 done for $^{84}Kr$ (panels a, b) and $^{131}Xe$ (c,d)
targets.}}
\end{figure}
Easily to see that they describe data on quantitative level.
Minimum $\hat{\chi}^2 = 0.54$ is obtained for $\sigma^{str}(10)$ and NDF 2
from Ref.~\cite{akopov7} at values of parameters equal
$\sigma_{q} = 2.34 \pm 0.14$ mb and $c = 0.088 \pm 0.01$.
Comparison with the Tables 1 and 2 from Ref.~\cite{akopov7} shows
that values of parameters and $\hat{\chi}^2$ in case of symmetric Lund
distribution function are close to those for the case of standard Lund
distribution function. Figure 1 shows that the good agreement of the leading
hadron approximation with the data is accidental, because the agreement
worsens with increasing of $N_{had}$ from 1 to the more realistic values for
multiplicity of hadrons. Comparison with the one dimensional data shows that
the $R_{M}^{h}$ calculated with the parameters obtained in result of fit with
symmetric Lund distribution function visually does not differ from others. A
small difference there is in case of two dimensional data which are presented
in Figs. 2 and 3. Experimental points are the two dimensional data (filled
symbols) from Ref.~\cite{airapet1}, and the one dimensional data (open symbols)
from Ref.~\cite{airapet2}. The theoretical curves are the results of fit with
the standard Lund distribution function - solid curves; with leading hadron
distribution function - dashed curves; with the symmetric Lund distribution
function - dotted curves.
\section{Conclusions.}
\hspace*{1em}
The HERMES data~\cite{airapet1} were used to perform the fit for ITSM with
symmetric Lund distribution function. It was shown that using in the fit
the distribution of the constituent formation lengths of hadrons presented in
the form of finite sum gives the stable result beginning with $n \ge 5$. For
the final fit we used the value $n=10$. Two-parameter's fit gave satisfactory
agreement with data (see Table 1 and Figs. 2 and 3). Comparison with other
versions of distribution function used in previous fit showed that the results
in the cases of the symmetric and standard Lund distribution functions are close
enough.


\begin{thebibliography}{99}
\bibitem{ashman} J.Ashman et al., Z.Phys. {\bf C52}( 1991) 1
\bibitem{akopov2} N.Akopov, L.Grigoryan, Z.Akopov, Eur.Phys.J. {\bf C44} (2005) 219
\bibitem{akopov7} N.Akopov, L.Grigoryan, Z.Akopov, Eur.Phys.J. {\bf C70} (2010) 5
\bibitem{airapet1} A.Airapetian et al.,  Nucl.Phys. {\bf B780} (2007) 1;
\bibitem{andersson} B.Andersson et al., Phys.Rep. {\bf 97} (1983) 31
\bibitem{sjostrand} T.Sjostrand, L.Lonnblad, S.Mrenna, hep-ph/0108264 (2001);
LU TP 01-21
\bibitem{bialas2} A.Bialas, M.Gyulassy, Nucl.Phys. {\bf B291} (1987) 793;
T.Chmaj, Acta Phys.Pol. {\bf B18} (1987) 1131
\bibitem{airapet2} A.Airapetian et al.,  Phys. Lett. {\bf B577} (2003) 37-46;
\end{thebibliography}
\end{document}